\documentclass[superscriptaddress,aps,pra, twocolumn, showpacs,nofootinbib,longbibliography,notitlepage]{revtex4-2}
\usepackage{etex}
\usepackage{Physics}
\usepackage{soul}
\usepackage{amsmath,amssymb,amsthm,amstext,mathrsfs, mathtools}
\usepackage[colorlinks=true,citecolor=blue,urlcolor=blue]{hyperref}
\usepackage[pdftex]{graphicx}
\usepackage{times,txfonts}
\usepackage{braket}
\usepackage{color}
\usepackage{natbib}
\usepackage{amsmath,blkarray}
\usepackage{mathtools}
\usepackage{latexsym}
\usepackage{tabularx, booktabs}
\usepackage{graphics,epstopdf}
\usepackage{pgfplots}
\pgfplotsset{width=8 cm,compat=1.8}
\usepackage{graphicx}
\usepackage{float}
\usepackage{graphicx}
\usepackage{amsfonts}
\usepackage{subcaption}
\usepackage{enumerate}
\usepackage{color,soul}

\newcommand{\one}{\mathbb{I}}

\begin{document}
\title{Incompatibility assisted Zeno-like confinement enables unbounded sharing of nonlocality}
\author{Prerna Rao}
\email{prerna.rao@research.iiit.ac.in}
\affiliation{Centre for Quantum Science and Technology, International Institute of Information Technology, Hyderabad, Gachibowli, Telangana 500032, India}
\affiliation{Center for Computational Natural Sciences and Bioinformatics, International Institute of Information Technology, Hyderabad, Gachibowli, Telangana 500032, India}
\author{Som Kanjilal}
\email{somkanjilal@gmail.com}
\affiliation{International Iberian Nanotechnology Laboratory, Avenida Mestre Jos\'{e} Veiga, 4715-330 Braga, Portugal}

\begin{abstract}
To enable sequential sharing of Bell-nonlocality by an unbounded number of copies, each copy must satisfy two requirements. First, the measurements must be incompatible to extract nonlocality. Second, the post-processsed state, obtained as an equal mixture of the two post-measurement states, must remain nonlocal. We show that the incompatibility requirements impose nontrivial constraints on the choice of the initial nonlocal state and the amount of measurement noise required to ensure that the post-processed state remains nonlocal for an unbounded number of copies. We establish this result for two measurement scenarios, namely, when each copy performs unsharp measurements corresponding to a pair of anti-commuting Pauli observables, and when each copy performs probabilistic projective measurements (PPMs) of a pair of anti-commuting Pauli observables. Furthermore, we show that, in the asymptotic limit, the nonlocal post-processed states of all the copies are almost identical, leading to a quantum Zeno-like confinement within the nonlocal region.
\end{abstract}
\maketitle

\section{Introduction}
A unilateral nonlocality sharing scenario involves one Alice and multiple sequential Bobs\cite{Silva2015,Steffinlongo2022,Brown2020,sasmal2023a,ZJC26}. Initially, Alice and the first Bob share a bipartite nonlocal state. Each Bob performs a pair of suitably chosen incompatible measurements, records the measurement outcomes, and transfers the post-processed state, obtained as an equal mixture of two post-measurement states, to the next Bob. Meanwhile, Alice projectively measures a predetermined non-commuting (incompatible) pair and records the corresponding outcomes. The objective of the scenario is to find suitable measurement strategies such that the Bell-Clauser-Horne-Shimony-Holt (CHSH) inequality \cite{Clauser1969} is violated for every Alice-Bob pair. 

In this scenario, each Bob extracts nonlocality from the incoming state by performing a Bell test. Each such test disturbs the shared quantum state, thereby reducing the available quantum correlations. Unbounded sharing is achieved when the post-processed state remains confined within the nonlocal region even after an asymptotically large number of Bell tests. This phenomenon is analogous to the quantum Zeno effect \cite{Misrasudarshan}, in which an initially excited quantum state remains in its excited state when we repeatedly ask whether it has decayed by performing a suitable two-outcome projective measurement. In the present scenario, instead of repeatedly probing if a state has decayed, we repeatedly probe whether it remains nonlocal. Moreover, instead of repeatedly applying projective measurements, the evolution is generated by a sequential application of a quantum channel consisting of  incompatible measurements. Unbounded sharing of nonlocality occurs when the state is confined within the nonlocal region even after an arbitrarily large number of Bell tests.

Existing works \cite{Brown2020,Steffinlongo2022,sasmal2023a} focus on building strategies for Alice-Bob pairs that enable unbounded sharing by computing the associated Bell values . In this work, we study how nonlocality is dynamically preserved in the post-processed state when each Bob extracts information about it by performing incompatible measurements. We further show that, for the measurement strategies considered here, the post-processed state remains confined within the nonlocal region through a quantum Zeno-like mechanism, whereby the successive nonlocal post-processed states become arbitrarily close to one another\cite{peres80} as the number of Bell tests becomes large.

Any measurement must ultimately be implemented through physical operations, that inevitably introduce disturbance and irreversibility \cite{Maccone_2006,CL+12,BH+14} while extracting information about the quantum state. In the context of sequential sharing of nonlocal correlations, this principle presents a fundamental tension: each Bob extracts information about nonlocality via incompatible measurements, but in doing so necessarily disturbs the shared quantum state, thus limiting the amount of nonlocality preserved in the post-processed state. In particular, one may consider two classes of measurement strategies for each Bob: one in which both measurements are equally unsharp, and another in which one measurement is projective while the other is unsharp. The incompatibility argument shows that the latter strategy allows the measurements to remain incompatible while simultaneously minimizing the disturbance \cite{sasmal2023a}, making it suitable for unbounded sharing of unilateral nonlocality. The state disturbance induced by the measurement incompatibility required to demonstrate nonlocality has also been shown to be important for tirparite nonlocality sharing \cite{Futurework2026}.

This motivates the question of how repeated application of a measurement channel composed of incompatible measurements (beyond the two cases discussed in \cite{sasmal2023a}) transforms a quantum state while preserving its nonlocality. Each Bob performs a pair of incompatible positive operator-valued measures (POVMs) on one part of the shared state to test for violation of the Bell–CHSH inequality. In the sequential sharing scenario, each Bob must additionally ensure that the resulting post-processed state remains nonlocal so that subsequent Bobs can also observe Bell violations. We show that, in the asymptotic limit, for each Bob, the incompatibility requirement to observe Bell-CHSH violation together with the condition for preserving nonlocality of the post-processed state, implies that only a class of initial states and measurement strategies can yield unbounded sharing of unilateral nonlocality. We establish this result for two measurement scenarios: $(a)$ each Bob performs a pair of unsharp measurement and $(b)$ each Bob performs a pair of local-randomness-assisted sharp measurement (also known as the probabilistic projective measurement (PPM) ). In particular, the strategies discussed in \cite{Brown2020,sasmal2023a} arise as special cases of the broader class of strategies that are shown to be suitable for unbounded sharing.

This paper is organized as follows: In section~\ref{sec:postprocessed}, we analyze the structure of the post-processed state by treating each Bob’s operation as a quantum channel composed of two measurements. In Section~\ref{sec:Horodecki}, we study the structure of the nonlocality preservation condition for the sequential sharing scenario considered here, namely that the Horodecki function of the post-processed state exceeds unity\cite{Horodecki1995}. In Section~\ref{sec:incompatibility}, we show that, in the asymptotic limit, the simultaneous satisfaction of the incompatibility requirement for the violation of the Bell inequality and the nonlocality preservation condition imposes nontrivial constraints on the state and measurement parameters. In Section~\ref{sec:unboundedsharing}, we construct explicit projective-unsharp and projective-PPM protocols based on these constraints and demonstrate unbounded sharing of unilateral nonlocality for a mixed state. Finally, in Section~\ref{sec:outlook}, we discuss how this scenario can be understood as a quantum Zeno-like confinement of the post-processed state within the nonlocal region, and we conclude by outlining some future directions.

\section{Background: The quantum channel perspective of sequential Nonlocality sharing}
\label{sec:postprocessed}

The objective of unbounded sharing of unilateral nonlocality is to satisfy the following condition\cite{Bell1964} for an arbitrarily large value of $k$: 
\begin{equation}
\label{bell}
Tr[(A_{0}\otimes \mathcal{B}^{k}_{0}+A_{0}\otimes \mathcal{B}^{k}_{1}+ A_{1}\otimes \mathcal{B}^{k}_{0}-A_{1}\otimes \mathcal{B}^{k}_{1})\rho_{k-1}]>2
\end{equation}
where $\mathcal{B}_{j}^{k}$ and $A_{i}$ are the two-outcome POVM and projective measurements of the $k$th Bob and Alice, respectively. $\rho_{k-1}$ is the post-processed state of the $(k-1)$th Bob. The measurement performed by the $k$th Bob can be viewed as a local channel $\Lambda_k(*)$ acting on the post-processed state of the $(k-1)$th Bob, $\rho_{k-1}$. Taking the initial nonlocal state to be $\rho_0$, the post-processed state of the $k$th Bob can be written as
\begin{align}
\label{inductivefrk}
\rho_{k} &= \Lambda_{k}(\Lambda_{k-1}(....(\Lambda_{1}(\rho_{0}))...)). 
\end{align}

For $j\in\mathbb{Z}_2:=\{0,1\}$, the $j$th measurement of the $k$th Bob is a POVM measurement $\mathcal{B}_{j}^{k}$ of a Pauli operator $B^{k}_{j}$ given by: 
 \begin{equation}
\label{eq:noisy_obs}
\mathcal{B}_{j}^{k} = b_{j,k}\one+s_{j,k}B^{k}_{j},
\end{equation}
where $\one$ is the identity operator corresponding to white noise, $-1\leq b_{j,k}\leq 1$ and $s_{j,k}>0$ are the bias and strength parameters of the measurement, respectively \cite{Cheng2021}.  If $b_{j,k}=0$ and $s_{j,k}=1$, then the measurement reduces to the projective measurement of the Pauli operator $B_{j}^{k}$. We assume that the Pauli operators for each Bob are identical (non-adaptive), i.e, $B_{j}^{k}=B_{j}=\Pi_{0|j}-\Pi_{1|j}$ and anti-commuting, i.e., $B_0=\sigma_{0}$ and $B_1=\sigma_{1}$. The pair $(b_{j,k},s_{j,k})$ contains all the information about the noise. The POVM elements for $\mathcal{B}_{j}^{k}=\mathcal{B}^{k}_{0|j}-\mathcal{B}^{k}_{1|j}$ are given by 
\begin{equation}
\label{bobkpovm}
\mathcal{B}^{k}_{b|j}= \frac{(1 +(-1)^b r_{0|j,k})}{2}\Pi_{0|j}+\frac{(1+(-1)^b r_{1|j,k})}{2}\Pi_{1|j},    
\end{equation}
where $r_{\beta|j,k}=b_{j,k} +(-1)^\beta s_{j,k}$ with $\beta\in\mathbb{Z}_2$. The positivity condition of the POVM elements gives $|b_{j,k}|+s_{j,k}\leq 1$ \cite{Busch1996book}.

Let the post-measurement state for the $j$th measurement of the $k$th Bob be given by $\rho_{k|j}$. The post-processed quantum state of the $k$th Bob is then given by
\begin{equation}
\label{postmeasfrnonselective}
\rho_k = \frac{1}{2}(\rho_{k|0}+\rho_{k|1}).
\end{equation}

Among the many possible Kraus decompositions that realize the POVM given by Eq.\eqref{bobkpovm}, we consider two measurement schemes:
\begin{itemize}
    \item The first is the L\"uders measurement scheme \cite{Busch2016book}, with Kraus operators given by
\begin{equation}
\label{Krausforjk}
\sqrt{\mathcal{B}^{k}_{b|j}} = \sqrt{\frac{(1+(-1)^b r_{0|j,k})}{2}}\Pi_{0|j}+\sqrt{\frac{(1 +(-1)^b r_{1|j,k})}{2}}\Pi_{1|j}.
\end{equation}
For this measurement scheme, the post-measurement state for the $j$th measurement of the $k$th Bob is given by
\begin{equation}
\label{postmeastate}
\rho_{k|j}=\sum_{b=0}^{1}\sqrt{\mathcal{B}^{k}_{b|j}}\rho_{k-1}\sqrt{\mathcal{B}^{k}_{b|j}}.
\end{equation}
Using Eqs.~(\ref{Krausforjk}) and \eqref{postmeastate}, $\rho_{k|j}$ can be written as (see Appendix \ref{postmeasurementstate})  
\begin{equation}
\label{postmeaswithKraus}
\rho_{k|j} = \frac{1}{2}[(1+q_{j,k})\rho_{k-1}+(1-q_{j,k})(\one\otimes\sigma_{j})\rho_{k-1}(\one\otimes\sigma_{j})],
\end{equation}
where 
\begin{equation}
\label{reversibility}
q_{j,k}=\frac{1}{2}\left(\sqrt{(1+b_{j,k})^{2}-s_{j,k}^{2}}+\sqrt{(1-b_{j,k})^{2}-s_{j,k}^{2}}\right)
\end{equation}
is known as the reversibility parameter \cite{Cheng2021}(``quality factor" in \cite{Silva2015}) of the $j$th measurement of the $k$th Bob. Using Eq.~(\ref{postmeaswithKraus}) and Eq.~(\ref{postmeasfrnonselective}), the post-processed state for the $k$th Bob, $\rho_{k}$, can be written as
\begin{align}
\label{postmeasfork}
\rho_{k} & = \frac{1}{4}[(2+q_{0,k}+q_{1,k})\rho_{k-1}+(1-q_{0,k})(\one\otimes \sigma_{0})\rho_{k-1}(\one\otimes \sigma_{0})\nonumber\\
& +(1-q_{1,k})(\one\otimes \sigma_{1})\rho_{k-1}(\one\otimes \sigma_{1})].
\end{align}
Later, we assume that the measurements are unsharp, i.e. $b_{j,k}=0$ and $q_{j,k}=\sqrt{1-s_{j,k}^{2}}$, where the strength parameter $s_{j,k}$ is also known as unsharpness parameter.
\item The second is the recently introduced probabilistic projective measurement (PPM) scheme \cite{sasmal2023a}, whose POVM effects correspond to the choice $b_{j,k}=1-s_{j,k}$. For the PPM-PPM measurement scheme, we denote the strength parameters by
$s'_{j,k}$. The corresponding Kraus operators are given as
\begin{align}
\label{eq:ppmKraus1}
\mathcal{K}_{1|j}^{k} & = \sqrt{1-s'_{j,k}}\one,\\
\label{eq:ppmKraus2}
\mathcal{K}_{2|j}^{k} & = \sqrt{s'_{j,k}}\Pi_{0|j},\\
\label{eq:ppmKraus3}
\mathcal{K}_{3|j}^{k} & = \sqrt{s'_{j,k}}\Pi_{1|j}.
\end{align}
The $k$th Bob performs a three-outcome POVM measurement of the Pauli operator $\sigma_{j}$ with Kraus operators $\{\mathcal{K}_{1|j}^{k}, \mathcal{K}_{2|j}^{k}, \mathcal{K}_{3|j}^{k}\}$, assigning the value $1$ to outcome $3$ and the value $0$ to the remaining two outcomes. For this measurement scheme, the post-measurement state for the $j$th measurement of the $k$th Bob is given by
\begin{equation}
\label{ppmpostmeastate}
\rho'_{k|j}=\sum_{b=1}^{3}\mathcal{K}^{k}_{b|j}\rho_{k-1}\mathcal{K}^{k}_{b|j}.
\end{equation}
Substituting Eqs.~(\ref{eq:ppmKraus1})--(\ref{eq:ppmKraus3}) into Eq.~(\ref{ppmpostmeastate}), we obtain  
\begin{equation}
\label{ppmpostmeaswithKraus}
\rho'_{k|j} = (1-\frac{s'_{j,k}}{2})\rho_{k-1}+\frac{s'_{j,k}}{2}(\one\otimes \sigma_{j})\rho_{k-1}(\one\otimes\sigma_{j}),
\end{equation}
Using Eq.~(\ref{ppmpostmeaswithKraus}) and Eq.~(\ref{postmeasfrnonselective}), the post-processed state of the $k$th Bob, $\rho'_{k}$,  for PPM-PPM measurement strategy can be written as
\begin{align}
\label{ppmpostmeasfork}
\rho'_{k} & = \frac{1}{4}[(4-s'_{0,k}-s'_{1,k})\rho_{k-1}+s'_{0,k}(\one\otimes \sigma_{0})\rho_{k-1}(\one\otimes \sigma_{0})\nonumber\\
& +s'_{1,k}(\one\otimes \sigma_{1})\rho_{k-1}(\one\otimes \sigma_{1})].
\end{align}
\end{itemize}
Note that both Eqs.~\eqref{postmeasfork} and \eqref{ppmpostmeasfork} are convex combinations of three quantum states, namely, $\rho_{k-1}$, $(\one\otimes \sigma_{0})\rho_{k-1}(\one\otimes \sigma_{0})$, and $(\one\otimes \sigma_{1})\rho_{k-1}(\one\otimes \sigma_{1})$. Thus, for both schemes, the post-processed state for the $k$th bob (denoted generically by $\rho_k$) can be expressed as
\begin{equation}
\label{eq:genpostprocessfork}
\rho_{k} = a_{k}\rho_{k-1} +b_{k}(\one\otimes\sigma_{0})\rho_{k-1}(\one\otimes\sigma_{0}) +c_{k}(\one\otimes\sigma_{1})\rho_{k-1}(\one\otimes\sigma_{1})
\end{equation}
where $a_k,b_k,c_k\ge0$ and
$a_k+b_k+c_k=1$.
Furthermore, comparing Eqs.~\eqref{postmeasfork} and \eqref{ppmpostmeasfork}, we observe that the post-processed state $\rho_k$ corresponding to a pair of L\"uders measurements with reversibility parameters $q_{0,k}$ and $q_{1,k}$, is identical to the post-processed state $\rho'_k$ corresponding to a PPM scheme with parameters $s'_{0,k}=1-q_{0,k}$ and $s'_{1,k}=1-q_{1,k}$. Thus, the two schemes generate identical post-processed states and identical state trajectories under the identification $s'_{j,k}=1-q_{j,k}$. The distinction between the schemes arises from the measurement statistics.

Using the expressions for the post-processed states derived in this section, we now characterize the condition for preservation of nonlocality of the $k$th Bob's post-processed state, given an initial bipartite qubit nonlocal state, by considering the associated Horodecki function \cite{Horodecki1995}.

\section{Horodecki Criterion for the Post-Processed State After the kth Bob}
\label{sec:Horodecki}
 A necessary and sufficient condition for the post-processed state $\rho_k$ to violate the Bell--CHSH inequality is that its Horodecki function \cite{Horodecki1995} exceeds unity. We consider the case where the initial state is a bipartite qubit state given by:
 \begin{equation}
\label{initbelldia}
\rho_{0}=\frac{1}{4}(\one\otimes\one+\sum_{j=0}^{2}m_{j}\sigma_{j}\otimes\one+\sum_{j=0}^{2}n_{j}\one\otimes\sigma_{j}+\sum_{j=0}^{2}t_{j}\sigma_{j}\otimes\sigma_j)
 \end{equation}

where $m_j=\mathrm{Tr}[\rho_0(\sigma_j\otimes\one)]$, $n_j=\mathrm{Tr}[\rho_0(\one\otimes\sigma_j)]$
are the components of the local Bloch vectors of Alice and Bob, respectively, and $|t_0|\geq |t_1|\geq |t_2|$. Let $T$ be the $3\times 3$ diagonal correlation matrix with correlation coefficients $t_{m} = \mathrm{Tr}[\rho\, \sigma_m \otimes \sigma_m]$. 
 The quantum state $\rho_{0}$ is nonlocal, or equivalently, it violates the Bell-CHSH inequality, if and only if\cite{Horodecki1995}
\begin{equation}
\label{horo1}
H_{0}=t_{0}^{2}+t_{1}^{2}>1.
\end{equation}

Let us define the post-processed state after the $k$th Bob, $\rho_{k}$, as 
\begin{equation}
\label{eq:postprock}
\rho_{k} = \frac{1}{4}(\one\otimes\one+\sum_{j=0}^{2}m_{j}^{k}\sigma_{j}\otimes\one+\sum_{j=0}^{2}n_{j}^{k}\one\otimes\sigma_{j}+\sum_{j=0}^{2}t_{j}^{k}\sigma_{j}\otimes\sigma_j).
\end{equation}

If each Bob performs a pair of L\"uders measurements characterized by the reversibility parameters $\{q_{0,i},q_{1,i}\}_{i=1}^{k}$, then the scalar coefficients of Eq.~\eqref{eq:postprock} can be written as (see Appendix \ref{horodeckiproof})
\begin{align}
\label{localalice}
m_{j}^{k} & = m_{j}, \quad n_{0}^{k}=
\frac{n_{0}}{2^{k}}
\prod_{m=1}^{k}(1+q_{1,m}),
\\
\label{localcoeff1sss}
n_{1}^{k}
&=
\frac{n_{1}}{2^{k}}
\prod_{m=1}^{k}(1+q_{0,m}),
\quad
n_{2}^{k}=\frac{n_{2}}{2^{k}}\prod_{m=1}^{k}(q_{0,m}+q_{1,m}),\\
\label{0corrcoefffrks}
t_{0}^{k} & = \frac{t_{0}}{2^{k}}\prod_{i=1}^{k}(1+q_{1,i}),\quad
t_{1}^{k}  = \frac{t_{1}}{2^{k}}\prod_{i=1}^{k}(1+q_{0,i}),\\
\label{2corrcoefffrks}
t_{2}^{k} & = \frac{t_{2}}{2^{k}}\prod_{i=1}^{k}(q_{0,i}+q_{1,i}).
\end{align}
Since the reversibility parameters are positive and upper bounded by one, the two largest correlation coefficients of $\rho_{k}$ are $t_{0}^{k}$ and $t_{1}^{k}$. The post-processed state for the $k$th Bob, $\rho_{k}$, is nonlocal if and only if \cite{Horodecki1995}
\begin{align}
\label{horode}
(t_{0}^{k})^{2}+(t_{1}^{k})^{2} & > 1,\\
\label{horo}
\underbrace{t_{0}^{2}\prod_{i=1}^{k}\qty(\frac{1+q_{1,i}}{2})^{2}+t_{1}^{2}\prod_{i=1}^{k}\qty(\frac{1+q_{0,i}}{2})^{2}}_{H(\{q_{1,i},q_{0,i}\}_{i=1}^{k})} & >1.
\end{align}
For convenience, we denote the Horodecki function for the $k$th Bob as $H_k$. Note that $\frac{(1+q_{j,i})}{2}\leq 1$ for all $i,j$ where equality only holds for $q_{j,i}=1$ (or equivalently, $s_{j,i}=0$). For strictly non-projective measurements ($0<q_{j,i}<1$), each multiplicative factor satisfies
$0<\frac{1+q_{j,i}}{2}<1$. Thus, the correlation terms are monotonically decreasing, with $|t_{1}^{k}|< |t_{1}^{k-1}|$ and $|t_{0}^{k}|< |t_{0}^{k-1}|$. It follows that the set $\{H_{i}\}_{i=1}^{k}$ is also a monotonically decreasing sequence, satisfying $H_{1}> H_{2}> H_{3}>\ldots> H_{k}$.

For unsharp measurements, characterized by $b_{j,k}=0$, and $0<s_{j,k}<1$, and for a fixed choice of the initial nonlocal state, we restrict attention to the class of non-adaptive strategies in which the Pauli operators are anti-commuting. Within this class, the choice of the unsharpness (strength) parameters can be classified as follows according to the relative ordering of the two reversibility parameters at each step:

\begin{itemize}
\item[(I)] The $i$th Bob chooses $s_{0,i}\le s_{1,i}$, or equivalently
$q_{0,i}\ge q_{1,i}$, for all $i\le k$.
\item[(II)] The $i$th Bob chooses $s_{0,i}\ge s_{1,i}$, or equivalently
$q_{0,i}\le q_{1,i}$, for all $i\le k$.
\item[(III)] The ordering between the two measurements is allowed to vary with $i$. That is, there exist indices $i$ for which $s_{0,i}\le s_{1,i}$ (equivalently $q_{0,i}\ge q_{1,i}$) and other indices for which $s_{0,i}\ge s_{1,i}$ (equivalently $q_{0,i}\le q_{1,i}$).
\end{itemize}

We refer to strategies of type (I) and (II) as pure strategies and to strategies of type (III) as mixed strategies. In Appendix~\ref{suboptimalmixture}, we show that for any mixed strategy with associated Horodecki function value $H_k$, there exists a pure strategy with Horodecki value $H_k'$ such that $H_k'\ge H_k$. Therefore, when maximizing the Horodecki function for a fixed pair of Pauli observables, it is sufficient to restrict attention to pure strategies. The value of the Horodecki function for a pure strategy can be expressed by collecting the larger and smaller reversibility parameters at each step. 

In particular, for a scenario satisfying strategy (II), we may take $q_{\max,i}=q_{1,i}$, or equivalently $s_{min,i}=s_{1,i}$, and $q_{\min,i}=q_{0,i}$, or equivalently $s_{max,i}=s_{0,i}$, where $q_{\max,i}=\sqrt{1-s_{\min,i}^2}$ corresponds to the less disturbing measurement and $q_{\min,i}=\sqrt{1-s_{\max,i}^2}$ corresponds to the more disturbing one. The nonlocality condition, expressed in terms of the Horodecki function, then reduces to
\begin{equation}
\label{nlpreser}
t_{0}^{2}\prod_{i=1}^{k}\left(\frac{1+q_{\max,i}}{2}\right)^2
+
t_{1}^{2}\prod_{i=1}^{k}\left(\frac{1+q_{\min,i}}{2}\right)^2 > 1.
\end{equation}
For the $k$th Bob, nonlocality is preserved provided that Eq.~(\ref{nlpreser}) is satisfied. For unbounded sharing, this condition must be satisfied for arbitrarily large values of $k$. 

Using the correspondence
$q_{j,i}=1-s'_{j,i}$,
the above analysis carries over directly to the case where each Bob performs PPM-PPM measurement scheme. The nonlocality preservation condition becomes
\begin{equation}
\label{ppmhoro_general}
\underbrace{t_{0}^{2}\prod_{i=1}^{k}\qty(1-\frac{s'_{1,i}}{2})^{2}+t_{1}^{2}\prod_{i=1}^{k}\qty(1-\frac{s'_{0,i}}{2})^{2}}_{H''(\{s'_{1,i},s'_{0,i}\}_{i=1}^{k})}  >1.
\end{equation}
Taking $f'_{j,i}=\qty(1-\frac{s'_{j,i}}{2})^{2}$, one can follow the procedure outlined in Appendix~\ref{suboptimalmixture} to show that $H''$ is maximized when $s'_{0,i}\geq s'_{1,i}$ for all $1\leq i\leq k$ or $s'_{0,i}\leq  s'_{1,i}$ for all $1\leq i\leq k$. Thus, the nonlocality preservation condition of the $k$th Bob for the PPM-PPM measurement scheme can be written as
\begin{equation}
\label{ppmhoro_pure}
t_{0}^{2}\prod_{i=1}^{k}\qty(1-\frac{s'_{min,i}}{2})^{2}+t_{1}^{2}\prod_{i=1}^{k}\qty(1-\frac{s'_{max,i}}{2})^{2} >1.
\end{equation}

For the sequential sharing scenario, the $k$th Bob tries to check whether the Bell-CHSH inequality is violated by performing a suitably chosen incompatible measurement pair in such a way that the post-processed state $\rho_{k}$ satisfies Eq.~(\ref{nlpreser}). Equipped with the nonlocality preservation conditions for both the unsharp-unsharp and the PPM-PPM measurement schemes, we now investigate how the incompatibility constraint restricts the admissible states and measurements for unbounded sharing.

\section{How Incompatibility Constraints Select the Feasible States and Measurements for Unbounded Sharing of Unilateral Nonlocality}
\label{sec:incompatibility}
We first consider the case where each Bob performs the unsharp-unsharp measurement scheme. From the perspective of each Bob, the extraction of nonlocality requires the measurements to be incompatible. For the choice of unsharp observables, the incompatibility requirement for each Bob is \cite{Busch2016book,Heinosaari2016,G_hne_2023}
\begin{align}
\label{incomptunsharpq}
q_{max,i}^2+q_{min,i}^2 & < 1,\\
\label{incomptunsharps}
s_{min,i}^2+s_{max,i}^2 & > 1.
\end{align}
Given an arbitrarily large integer $k$, ensuring that all $k$ Bobs can sequentially extract nonlocality in the scenario considered here requires identifying regions of $(s_{max,i}, s_{min,i})$ such that $H_i > 1$ and Eq.~(\ref{incomptunsharps}) is satisfied for all $i \leq k$.
Note that Eq.~(\ref{incomptunsharps}) implies $s^2_{max,i}>\frac{1}{2}$ or $q^2_{min,i}<\frac{1}{2}$. This ensures that
\begin{equation}
\label{loweboundqmin}
t_{1}^{2}\prod_{i=1}^{k}\qty(\frac{1+q_{min,i}}{2})^2< t_{1}^{2}\mathcal{C}^{2k}
\end{equation}
where 
\begin{equation}
\mathcal{C}=\frac{1}{2}\qty(1+\frac{1}{\sqrt{2}})\approx 0.85.    
\end{equation}
 Plugging Eq.~(\ref{loweboundqmin}) into Eq.~(\ref{nlpreser}) we get the following:
\begin{equation}
\label{ublbfrkk}
 t_{0}^{2}\prod_{i=1}^{k}\qty(\frac{1+q_{max,i}}{2})^2 > 1-t_{1}^{2}\prod_{i=1}^{k}\qty(\frac{1+q_{min,i}}{2})^2>1-t_{1}^{2}\mathcal{C}^{2k}.
\end{equation}
Using the fact that $q_{max,i}\leq1$ for any given $i$, we can derive an upper bound for the product term $t_{0}^{2}\prod_{i=1}^{k}\qty(\frac{1+q_{max,i}}{2})^2$ as follows:
\begin{equation}
\label{ublbfrk}
t_{0}^{2}\qty(\frac{1+q_{max,i}}{2})^2 >t_{0}^{2}\prod_{i=1}^{k}\qty(\frac{1+q_{max,i}}{2})^2>1-t_{1}^{2}\mathcal{C}^{2k}.
\end{equation}
 As $k$ increases, the interval allowed by Eq.~(\ref{ublbfrk}) shrinks towards unity, thereby imposing increasingly stringent constraints on both the state and measurement parameters. Since
$(1-t_{1}^{2}\mathcal C^{2k})\to1^{-}$ i.e., it converges to $1$ from below
as $k\to\infty,$ Eq.~(\ref{ublbfrk}) implies $t_{0}^{2}\prod_{i=1}^{k}
\left(\frac{1+q_{\max,i}}{2}\right)^2
\to1^{-}.$ Consequently, the upper bound $t_{0}^{2}\left(\frac{1+q_{\max,i}}{2}\right)^2$ must also approach unity. Since $t_{0}^{2}\leq 1$ and $
\frac{1+q_{\max,i}}{2}\leq 1$, this is possible only if $t_0^2=1$ and $q_{\max,i}\to 1^{-}$ for all $i$. Alternatively, if there exists an $i$ such that $q_{\max,i}$ is bounded away from $1$, then for sufficiently large $k$, the lower bound in Eq.~(\ref{ublbfrk}) cannot be satisfied, and hence the post-processed state cannot remain nonlocal.

 Thus, for unbounded sharing, the quantum states should satisfy $t_{0}^{2}=1$ and $q_{max,i}\to 1^{-}$ (equivalently, $s_{\min,i}\to 0^{+}$, i.e., the corresponding unsharp parameter approaches $0$ from the positive side) for all $i$. Therefore, satisfying the incompatibility constraint Eq.~\eqref{incomptunsharpq} requires   $q_{min,i}\to 0^{+}$ (i.e., $s_{\max,i}\to 1^{-}$).

We now discuss how incompatibility constraints select the suitable state and strength parameters for unbounded sharing in the case of the PPM-PPM measurement strategy. For a pair of PPM observables, it can be shown that the incompatibility requirement \cite{Busch2016book,Heinosaari2016,G_hne_2023} is given by
\begin{equation}
\label{incompppm}
s'_{\min,i}+s'_{\max,i}>1+(1-s'_{\min,i})(1-s'_{\max,i})>1.
\end{equation}
This, in turn, implies $s'_{\max,i}>\frac{1}{2}$. Proceeding along the same lines as in the case of the unsharp-unsharp measurement strategy, we obtain the following:
\begin{equation}
\label{ppmsharingfrk}
t_{0}^{2}\qty(1-\frac{s'_{min,i}}{2})^{2}\geq t_{0}^{2}\prod_{i=1}^{k}\qty(1-\frac{s'_{min,i}}{2})^{2}>1-t_{1}^{2}\qty(\frac{3}{4})^{2k}.
\end{equation}
The same argument as before shows that unbounded sharing for the PPM-PPM measurement occurs only if $t_0^2=1$ and $s'_{\min,i}\to 0^{+}$ for all $i$. The incompatibility constraint then requires $s'_{\max,i}\to 1^{-}$ for all $i$. 

Thus, for the measurement strategies considered here, a necessary condition for unbounded sharing of unilateral nonlocality is that $t_0^2=1$ and, for each Bob, either the first or the second measurement is almost projective, while the other has almost vanishing strength. This class includes the projective-unsharp (or projective-PPM) measurement strategies shown in \cite{Brown2020,sasmal2023a} to achieve unbounded sharing. These strategies, characterized by $t_0^2=1$, $s_{\min,i}\to 0^{+}$ ($s'_{\min,i}\to 0^{+}$), and $s_{\max,i}=1$ ($s'_{\max,i}=1$), constitute the simplest feasible choice for unbounded sharing of unilateral nonlocality.
The constraints $s_{\max,i}\to 1^{-}$ and $s_{\min,i}\to 0^{+}$ restrict the evolution of the post-measurement states given by Eq.~\eqref{postmeaswithKraus}. For each $i$, the post-measurement state corresponding to $s_{\max,i}$ is separable, since the corresponding measurement is almost projective. Subsequently, $s_{min,i}\to 0^{+}$ implies that the corresponding post-measurement state is almost identical to $\rho_{i-1}$. Later, we show that, for this choice of incompatible measurements, in the asymptotic limit, the distance between successive post-processed states converges to zero, yielding a Zeno-like "collapse" within the nonlocal regime.

In the next section, we explicitly construct an example of unbounded sharing using the conditions obtained in this section for both the projective-unsharp and the projective-PPM measurement strategies. 

\section{Strategies for unbounded sharing of unilateral nonlocality from an initial mixed state}
\label{sec:unboundedsharing}

We discuss the case where $t_{0}=1$ ($t_{0}=-1$ case will proceed similarly). Consider the initial state as follows,
\begin{align}
\label{mixedstate_theta}
\rho_{0} & =\frac{1}{4}\left(\one\otimes\one +\sigma_0\otimes\sigma_0 + t\sigma_1\otimes\sigma_1 - t\sigma_2\otimes\sigma_2 \right),\\
 \label{convexmix}
& = p\ket{\Psi^+}\bra{\Psi^+} + (1-p)\ket{\Phi^+}\bra{\Phi^+}
\end{align}
where
\begin{equation}
\label{pandt}
p\in[0,1]\ \text{and}\ t=2p-1.
\end{equation}
In Appendix \ref{initialstate} we show that this is the only possible choice for the initial Bell diagonal mixed quantum state with $t_{0}=1$.
Note that Eqs~\eqref{localalice}-\eqref{2corrcoefffrks} imply that $\rho_{k}$ remains Bell diagonal.
In accordance with the necessary condition derived in Sec.~\ref{sec:incompatibility}, we choose the projective--unsharp measurement strategy as 
\begin{equation}
\label{bobmixed}
\mathcal{B}_{0}^k = \sigma_0,
\qquad
\mathcal{B}_{1}^k = s_{min,k}\sigma_1,
\end{equation}
Alice's observables are parametrized as
\begin{align}
A_0 &= \cos\theta_0\sigma_0+\sin\theta_0\sigma_1, \quad A_1 = \cos\theta_1\sigma_0+\sin\theta_1\sigma_1.
\end{align}
Since \[\{A_{0},A_{1}\}=2\cos(\theta_{0}-\theta_1)\one,\] the observables are incompatible as long as $\theta_{0}\neq \frac{\pi}{2}+\theta_1$.
Using the Bell-diagonal form of the post-measurement state in Appendix \ref{horodeckiproof} with correlation coefficients $t_{0}^{k-1},t_{1}^{k-1}$ and $t_{2}^{k-1}$, each term of the Bell-CHSH functional of Eq.~\eqref{bell} can be written as follows
\begin{align}
\langle A_0\otimes\mathcal{B}_{0}^k\rangle &= t_0^{k-1}\cos\theta_0,\quad\langle A_0\otimes\mathcal{B}_{1}^k\rangle = s_{min,k} t_1^{k-1}\sin\theta_0,\\
\langle A_1\otimes\mathcal{B}_{0}^k\rangle &= t_0^{k-1}\cos\theta_1,\quad
\langle A_1\otimes\mathcal{B}_{1}^k\rangle = s_{min,k} t_1^{k-1}\sin\theta_1.
\end{align}
Substituting the above correlators into the Bell-CHSH expression in Eq.~(\ref{bell}), we obtain
\begin{equation}
\label{Ikmixedtheta}
I_k = t_0^{k-1}(\cos\theta_0+\cos\theta_1) + s_{min,k}t_1^{k-1}(\sin\theta_0-\sin\theta_1),
\end{equation}
where $t_{0}^{0}=t_{0}=1$ and $t_{1}^{0}=t_{1}=t$ are the correlation coefficients of the initial state.

Introducing the parametrization
\begin{equation}
\label{phideltamixed}
\phi=\frac{\theta_0+\theta_1}{2},
\qquad
\delta=\frac{\theta_0-\theta_1}{2},
\end{equation}
and using
\begin{align}
\cos\theta_0+\cos\theta_1 &= 2\cos\phi\cos\delta,\\
\sin\theta_0-\sin\theta_1 &= 2\cos\phi\sin\delta,
\end{align}
Eq.~(\ref{Ikmixedtheta}) becomes
\begin{equation}
\label{Ikmixedphi}
I_k =2\cos\phi\left[t_0^{k-1}\cos\delta+s_{min,k}t_1^{k-1}\sin\delta\right].
\end{equation}
$I_{k}$ in Eq.~\eqref{Ikmixedphi} is maximized for $\phi=0$, yielding
\begin{equation}
\label{Ikmixedfinal}
I_k =2\left[t_0^{k-1}\cos\delta + s_{min,k}t_1^{k-1}\sin\delta\right].
\end{equation}
Using Eqs.~\eqref{0corrcoefffrks} with $q_{0,i}=q_{min,i}=0$, $q_{1,i}=q_{max,i}=\sqrt{1-s_{min,i}^{2}}$, $t_{0}=1$ and $t_{1}=t$ in Eq.~\eqref{Ikmixedfinal} we get
\small
\begin{equation}
\label{Ikmixedrecursive}
I_k =2\left[\frac{\cos\delta}{2^{k-1}}\prod_{i=1}^{k-1}(1+\sqrt{1-s_{min,i}^{2}})+\frac{t s_{min,k}\sin\delta}{2^{k-1}}\right], \quad \text{if } k\geq 2,
\end{equation}
\normalsize
and
\begin{equation}
\label{I1mixedrecursive}
I_{1}=2\left[\cos\delta+t s_{min,1}\sin\delta\right].
\end{equation}
To obtain a violation of the Bell-CHSH inequality for the $k$th Bob, we require $I_k>2$ for all $k\geq 1$. Imposing the condition $I_k>2,$ and using Eqs.~(\ref{Ikmixedrecursive}) and (\ref{I1mixedrecursive}), we obtain
\begin{align}
\label{sconditionmixed}
s_{min,k}
&>
\frac{
2^{k-1}
-
\cos\delta
\prod_{i=1}^{k-1}(1+\sqrt{1-s_{min,i}^{2}})}{t\sin\delta},
\quad \text{if }k\geq 2,\\
\label{1conditionmixed}
s_{min,1}&> \frac{1-\cos\delta}{t\sin\delta}=\frac{1}{t}\tan\frac{\delta}{2}.
\end{align}
The lower bounds given by Eqs~\eqref{sconditionmixed} and \eqref{1conditionmixed} are positive as long as $\delta>0$. To show unbounded sharing, we need to show that it is possible to choose a sequence $\{s_{min,l}\}_{l=1}^{k}$ that satisfies this lower bound for an arbitrarily large $k$. In particular, we need to show that it is possible to choose $\{s_{min,l}\}_{l=1}^{k}$ that satisfy
\begin{align}
\label{feasiblesequence}
1>s_{min,l} &> \frac{2^{l-1}-\cos\delta\prod_{i=1}^{l-1}(1+\sqrt{1-s_{min,l}^{2}})}{t\sin\delta}
\quad \text{if }l\geq 2,\\
\label{initfeasse1}
1>s_{min,1} &> \frac{1}{t}\tan\frac{\delta}{2}
\end{align}
for $\delta>0$.
Consider the following sequence, defined for $\delta>0$, which satisfies Eq.~\eqref{sconditionmixed}:
\begin{equation}
\label{mixedrecursion}
s_{min,l}(\omega,\delta)  =
\begin{cases}
(1+\omega)\dfrac{
2^{l-1}-\cos\delta\prod_{i=1}^{l-1}(1+\sqrt{1-s_{min,i}^{2}})}{t\sin\delta},
& l\geq 2,\\[2ex]
\dfrac{(1+\omega)}{t}\tan\frac{\delta}{2},& l=1.
\end{cases}
\end{equation}
where $\omega > 0$.

To establish that $\{s_{\min,l}(\omega,\delta)\}_{l=1}^{k}$ is a valid sequence of unsharpness parameters, one must show that it satisfies $0 \leq s_{\min,l}(\omega,\delta) \leq 1$ for all $1\leq l\leq k$, given any arbitrary integer $k$. It can be shown that $\{s_{min,l}(\omega,\delta)\}_{l=1}^{k}$ forms a positive monotonically increasing sequence(see Appendix C of \cite{Brown2020}), i.e.,
\[0<s_{min,1}(\omega,\delta)<s_{min,2}(\omega,\delta)<\cdots<s_{min,k}(\omega,\delta)\]
 and the largest term converges to zero as $\delta\to 0^{+}$(see Appendix D of \cite{Brown2020}) i.e, 
\[\lim_{\delta\to 0^{+}}s_{min,k}(\omega,\delta)=0\]
Given an arbitrary large integer $k$, since each term of the monotonically increasing sequence $\{s_{min,l}\}_{l=1}^{k}$ is positive, and the largest term $s_{min,k}(\omega,\delta)$ converges to zero in the limit $\delta\to 0^{+}$, we can conclude that all the terms converge to zero from the positive side. Thus, there exist suitable values for $\delta$ and $\omega$ such that all the terms in the sequence $\{s_{min,l}\}_{l=1}^{k}$ lie between $0$ and $1$. 
Consequently, $\{s_{min,l}(\omega,\delta)\}_{l=1}^{k}$ is a feasible sequence for unbounded sharing of unilateral nonlocality in the projective-unsharp scenario.

For the case where each Bob employs the Projective-PPM measurement strategy, the proof proceeds along the same lines, as the initial state is Bell diagonal and it gives identical results. We choose the effective observables as
\begin{equation}
\label{bobmixedppm}
\mathcal{B}_{0}^{k}=\sigma_0,
\qquad
\mathcal{B}_{1}^{k}=(1-s'_{min,k})\one+s'_{min,k}\sigma_1.
\end{equation}
For the projective-PPM strategy, $\langle A_{j}\otimes\mathcal{B}_{0}^{k}\rangle$ are identical to those obtained by the projective-unsharp strategy.  In general, only the correlators involving the second observable $\mathcal{B}_{1}^{k}$, i.e, $\langle A_{j}\otimes\mathcal{B}_{1}^{k}\rangle$ will differ from those of the projective-unsharp strategy and are given by
\begin{align}
\langle A_{j}\otimes \mathcal{B}_{1}^{k}\rangle
&=
(1-s'_{min,k})\langle A_{j}\otimes\one\rangle
+s'_{min,k}\langle A_{j}\otimes\sigma_1\rangle.
\end{align}
Since the initial state in Eq.~(\ref{mixedstate_theta}) is Bell-diagonal, Eq.~\eqref{localalice}-\eqref{localcoeff1sss} imply that the local Bloch vectors vanish for $k$, so that
\begin{equation}
\label{localvanish}
\langle A_{j}\otimes\one\rangle=0
\end{equation}
for all $j$ and for all post-processed states $\rho_{k}$. Hence, for the family of nonlocal Bell-diagonal states considered here, the Bell-CHSH expression for the projective-PPM strategy is identical to that of the projective-unsharp strategy. Thus, Eqs.~(\ref{Ikmixedtheta})--(\ref{mixedrecursion}) remain valid for the projective-PPM scheme. The Bell expression is again given by Eq.~(\ref{Ikmixedfinal}), and the same feasible sequence of measurement parameters establishes unbounded sharing. 

\section{Outlook and Discussion}
\label{sec:outlook}
In summary, we showed that if each Bob employs either a pair of unsharp POVM measurements or of PPM measurements whose projective counterparts are identical anti-commuting Pauli operators, then an arbitrarily large number of Bobs can sequentially extract nonlocality by violating the Bell-CHSH inequality only if, for every Bob, one of the measurements is almost-projective, the other is very weak, and one of the correlation coefficients of the initial state has unit magnitude. Under the assumptions of non-adaptivity and anti-commuting Pauli operators, the results of \cite{Brown2020,sasmal2023a} establish the sufficiency of the projective-unsharp and projective-PPM measurement strategies for unbounded sharing. Our results complement these works by showing that, under the same assumptions, these strategies constitute  the simplest feasible strategies for unbounded sharing. Furthermore, we characterize the class of initial states that allow unbounded sharing of unilateral nonlocality. \\
This situation is analogous to the quantum Zeno effect \cite{Misrasudarshan}, with a quantum channel playing the role typically associated with the intervening measurement \cite{Misrasudarshan,FacchiPascazio2002}. In its most basic form \cite{Misrasudarshan}, the Quantum Zeno effect refers to the phenomenon in which a quantum system in an excited energy state does not decay to the ground state when it is repeatedly probed to check if the decay has occurred. Further generalizations \cite{FacchiPascazio2002,FacchiPascazio2008} show that if one considers the system prepared in an initial state, repeatedly performs a fixed measurement, and at each step considers the coarse-grained state by summing over the post-measurement states, then the resulting state at each step remains confined within the orthogonal measurement subspaces, and the transition probabilities between the subspaces vanish in the asymptotic limit of iterations. This mechanism is used to confine the quantum system within a subspace that preserves a desired property by repeatedly intervening in the evolution, for example, by protecting the entanglement from decoherence \cite{SMFF+08}, suppressing loss in a Bose-Einstein condensate \cite{Schtzhold_2010} or in a driven-dissipative system \cite{Secl2022}, or confining the quantum state within the error-free subspace \cite{V96}. In the present scenario, the desired property is nonlocality and the intervention is a Bell test. Given an initial nonlocal state, each Bob asks whether the incoming state (i.e. the post-processed state of the previous Bob) is nonlocal or not by performing suitably chosen incompatible measurements. After each Bell test, the post-processed state retains a smaller amount of entanglement. Typically, the entanglement decreases progressively until the state becomes local after a finite number of Bobs. In contrast, for unbounded sharing, the measurements performed by each Bob are chosen so that the post-processed state never becomes local, remaining confined within the nonlocal region even after an arbitrarily large number of Bell tests. To see this more clearly, consider the distance between the post-processed states $\rho_k$ and $\rho_{k-1}$ when the initial state is the Bell diaognal state given by Eq.~\eqref{mixedstate_theta}. Using Eqs.~\eqref{localalice}-\eqref{2corrcoefffrks} with $t_{0}=1,t_{1}=t_{2}=t$ and $n_{0}=n_{1}=n_{2}=m_{0}=m_{1}=m_{2}=0$ we obtain
\begin{align}
\label{succdist}
\Delta_{k}^{2}&=\|\rho_{k-1}-\rho_k\|^2
=
\frac{1}{4}
\Bigg[
\left(\frac{1-q_{1,k}}{2}\right)^2
\prod_{j=1}^{k-1}
\left(\frac{1+q_{1,j}}{2}\right)^2
\nonumber\\
&\qquad+
t^{2}
\left(\frac{1-q_{0,k}}{2}\right)^2
\prod_{j=1}^{k-1}
\left(\frac{1+q_{0,j}}{2}\right)^2
\nonumber\\
&\qquad+
t^{2}
\left(1-\frac{q_{1,k}+q_{0,k}}{2}\right)^2
\prod_{j=1}^{k-1}
\left(\frac{q_{0,j}+q_{1,j}}{2}\right)^2
\Bigg],
\end{align}
where $\|X\|^{2}=\mathrm{Tr}(X^\dagger X).$
Now consider the strategy characterized by
$q_{1,j}\to 1^{-}, q_{0,j}\to 0^{+}$,
for all $j\leq k$. We analyze each term in Eq.~\eqref{succdist} separately.
For the first term, $\left(\frac{1-q_{1,k}}{2}\right)^2\prod_{j=1}^{k-1}
\left(\frac{1+q_{1,j}}{2}\right)^2$, the product $\prod_{j=1}^{k-1}\left(\frac{1+q_{1,j}}{2}\right)^2$ remains bounded since each factor is at most unity. Since $\lim_{q_{1,k}\to 1^{-}}(1-q_{1,k})=0$, it follows that the first term vanishes for any finite $k$.
For the second term, using $q_{0,j}\to0^{+}$, we have
$\left(\frac{1-q_{0,k}}{2}\right)^{2}
\prod_{j=1}^{k-1}
\left(\frac{1+q_{0,j}}{2}\right)^{2}
\sim
\frac{1}{4^k}$.
Finally, for the third term, since $q_{1,j}\to1, 
q_{0,j}\to0$, we have $
\frac{q_{0,j}+q_{1,j}}{2}\to\frac{1}{2}
$. Therefore, $\left(1-\frac{q_{1,k}+q_{0,k}}{2}\right)^2\prod_{j=1}^{k-1}
\left(\frac{q_{0,j}+q_{1,j}}{2}\right)^{2}\sim
\frac{1}{4^{k}}$.
Thus, for a finite $k$ we can write
\begin{equation}
\Delta_{k}^{2}\sim \frac{2t^{2}}{4^{k+1}}=\frac{t^{2}}{2^{2k-1}}.
\end{equation}
Hence, for arbitrarily large $k$, the feasible strategies for unbounded sharing identified in this work make the distance between successive post-processed states arbitrarily small, while maintaining incompatibility of the measurements. Furthermore, Eq.~\eqref{0corrcoefffrks} implies that $t_{0}^{k}\sim t_{0}=1$ and $t_{1}^{k}\sim \frac{t}{2^{k}}$ when $q_{1,j}\to 1^{-}, q_{0,j}\to 0^{+}$,
for all $j\leq k$, thus nonlocality of the postprocessed state is preserved for all finite $k$ since $H_{k}>1$.

Therefore, after a couple of steps, the successive nonlocal post-processed states become almost identical, thereby attaining a Zeno-like confinement to preserve nonlocality \cite{peres80}. However, it is important to emphasize that this confinement is inherently asymmetric: it relies on the presence of an almost projective measurement as an anchor to maintain incompatibility. It would be interesting to investigate whether similar Zeno-like confinement mechanisms arise in other sequential nonlocality-sharing scenarios \cite{skd26,ZJC26}.

\section*{Acknowledgments}
PR thanks the University Grants Commission, Government of India, for the Junior Research Fellowship. SK acknowledges the support from Digital Horizon Europe project, FoQaCia (\textit{Foundations of quantum computational advantage}), GA no 202070558, funded by the European Union and NSERC (Canada). SK also acknowledges fruitful discussions with Alok Kumar Pan, Sauradeep Sasmal and Debarshi Das.


\bibliography{references} 
\begin{widetext}
\appendix
\section{Derivation of Equation (\ref{postmeaswithKraus})}\label{postmeasurementstate}
Using the Kraus operators given in Eq.~(\ref{Krausforjk}), we obtain
\begin{align*}
\rho_{k|j}& = \sum_{b}(\one\otimes\sqrt{\mathcal{B}^{k}_{b|j}})\rho_{k-1}(\one\otimes\sqrt{\mathcal{B}^{k}_{b|j}}^{\dagger})\\
& = \sum_{b}\Bigg(\sqrt{\frac{(1+(-1)^b r_{0|j,k})}{2}}\one\otimes\Pi_{0|j}
+\sqrt{\frac{(1 +(-1)^b r_{1|j,k})}{2}}\one\otimes\Pi_{1|j}\Bigg)\\
& \rho_{k-1}
\Bigg(\sqrt{\frac{(1+(-1)^b r_{0|j,k})}{2}}\one\otimes\Pi_{0|j}
+\sqrt{\frac{(1 +(-1)^b r_{1|j,k})}{2}}\one\otimes\Pi_{1|j}\Bigg)\\
& = \sum_{b=0}^{1}\Bigg[\frac{(1+(-1)^b r_{0|j,k})}{2}\,
\one\otimes\Pi_{0|j}\rho_{k-1}\one\otimes\Pi_{0|j}\\
&\qquad\qquad\qquad
+\frac{(1+(-1)^b r_{1|j,k})}{2}\,
\one\otimes\Pi_{1|j}\rho_{k-1}\one\otimes\Pi_{1|j}\Bigg]\\
&\qquad +\Bigg(\frac{\sqrt{(1+r_{0|j,k})(1+r_{1|j,k})}}{2}
+\frac{\sqrt{(1-r_{0|j,k})(1-r_{1|j,k})}}{2}\Bigg)\\
&\qquad \times
\left(\one\otimes\Pi_{0|j}\rho_{k-1}\one\otimes\Pi_{1|j}
+\one\otimes\Pi_{1|j}\rho_{k-1}\one\otimes\Pi_{0|j}\right).
\end{align*}

Using $r_{\beta|j,k}=b_{j,k}+(-1)^\beta s_{j,k}$, the above expression simplifies to
\begin{align}
\rho_{k|j}
&=
\one\otimes\Pi_{0|j}\rho_{k-1}\one\otimes\Pi_{0|j}
+\one\otimes\Pi_{1|j}\rho_{k-1}\one\otimes\Pi_{1|j}
\nonumber\\
&\quad
+q_{j,k}
\left(
\one\otimes\Pi_{0|j}\rho_{k-1}\one\otimes\Pi_{1|j}
+\one\otimes\Pi_{1|j}\rho_{k-1}\one\otimes\Pi_{0|j}
\right),
\label{postmeasfrj}
\end{align}
where the coefficients multiplying the diagonal terms reduce to unity, while the coefficient of the off-diagonal terms reduces to $q_{j,k}$, and
\[
q_{j,k}
=
\frac{1}{2}
\left(
\sqrt{(1+b_{j,k})^{2}-s_{j,k}^{2}}
+
\sqrt{(1-b_{j,k})^{2}-s_{j,k}^{2}}
\right)
\]
is the reversibility parameter defined in Eq.~(\ref{reversibility}). Using
\begin{align*}
\rho_{k-1}
&=
\one\otimes\Pi_{0|j}\rho_{k-1}\one\otimes\Pi_{0|j}
+\one\otimes\Pi_{1|j}\rho_{k-1}\one\otimes\Pi_{1|j}\\
&\quad
+\one\otimes\Pi_{0|j}\rho_{k-1}\one\otimes\Pi_{1|j}
+\one\otimes\Pi_{1|j}\rho_{k-1}\one\otimes\Pi_{0|j},
\end{align*}
and
\begin{align*}
(\one\otimes B_j)\rho_{k-1}(\one\otimes B_j)
&=
\one\otimes\Pi_{0|j}\rho_{k-1}\one\otimes\Pi_{0|j}
+\one\otimes\Pi_{1|j}\rho_{k-1}\one\otimes\Pi_{1|j}\\
&\quad
-\one\otimes\Pi_{0|j}\rho_{k-1}\one\otimes\Pi_{1|j}
-\one\otimes\Pi_{1|j}\rho_{k-1}\one\otimes\Pi_{0|j},
\end{align*}
which follows from $B_j=\Pi_{0|j}-\Pi_{1|j}$ and $\Pi_{0|j}+\Pi_{1|j}=\one$. Hence, we obtain
\begin{equation}
\rho_{k|j}
=
\frac{1}{2}
\left[
(1+q_{j,k})\rho_{k-1}
+
(1-q_{j,k})
(\one\otimes B_j)\rho_{k-1}(\one\otimes B_j)
\right],
\end{equation}
which is Eq.~(\ref{postmeaswithKraus}).

\section{Proof of Eqs.~(\ref{localalice}-\ref{2corrcoefffrks})}
\label{horodeckiproof}

Since both the Lüders and PPM schemes generate the same channel representation given by Eq.~(\ref{eq:genpostprocessfork}), it suffices to analyze the action of the generic channel coefficients $\{a_i,b_i,c_i\}$.

We begin with a bipartite state $\rho_0$ and derive an expression for $\rho_k$ from Eq.~(\ref{inductivefrk}). To this end, we analyze the action of the channel $\Lambda_i$ on the state
\[
\rho_{i-1}
=
\frac{1}{4}
\left(
\one\otimes\one
+
\sum_{j=0}^{2}m_{j}^{i-1}\sigma_{j}\otimes\one
+
\sum_{j=0}^{2}n_{j}^{i-1}\one\otimes\sigma_{j}
+
\sum_{j=0}^{2}t_{j}^{i-1}\sigma_j\otimes\sigma_j
\right),
\]
and use the fact that $\{a_i,b_i,c_i\}$ forms a probability distribution to obtain
\begin{align}
\rho_i
&=
a_i\rho_{i-1}
+b_i(\one\otimes\sigma_0)\rho_{i-1}(\one\otimes\sigma_0)
+c_i(\one\otimes\sigma_1)\rho_{i-1}(\one\otimes\sigma_1)
\nonumber\\
&=
\frac{1}{4}
\Big[
\one\otimes\one
+
\sum_{j=0}^{2}m_{j}^{i-1}\sigma_{j}\otimes\one
+
n_{0}^{i-1}(a_i+b_i-c_i)\one\otimes\sigma_{0}
+
n_{1}^{i-1}(a_i-b_i+c_i)\one\otimes\sigma_{1}
\nonumber\\
&\qquad
+
n_{2}^{i-1}(a_i-b_i-c_i)\one\otimes\sigma_{2}
+
t_{0}^{i-1}(a_i+b_i-c_i)\sigma_{0}\otimes\sigma_{0}
+
t_{1}^{i-1}(a_i-b_i+c_i)\sigma_{1}\otimes\sigma_{1}
\nonumber\\
&\qquad
+
t_{2}^{i-1}(a_i-b_i-c_i)\sigma_{2}\otimes\sigma_{2}
\Big].
\label{inductivek2k-1}
\end{align}

In deriving the above expression, we use
\[
(\one\otimes\sigma_\alpha)
(\sigma_j\otimes\sigma_j)
(\one\otimes\sigma_\alpha)
=
\sigma_j\otimes
(\sigma_\alpha\sigma_j\sigma_\alpha),
\]
together with
\[
\sigma_0\sigma_j\sigma_0=-\sigma_j,
\qquad j\in\{1,2\},
\]
and
\[
\sigma_1\sigma_j\sigma_1=-\sigma_j,
\qquad j\in\{0,2\}.
\]

Since the channel acts only on Bob's subsystem, the local Bloch vector of Alice remains unchanged:
\begin{equation}
m_j^i=m_j^{i-1}=m_j,
\qquad j\in\{0,1,2\}.
\end{equation}

Consequently, we obtain
\begin{align}
\label{localcoeff0}
n_{0}^{i}
&=
n_{0}^{i-1}(a_i+b_i-c_i),
\\
\label{localcoeff1}
n_{1}^{i}
&=
n_{1}^{i-1}(a_i-b_i+c_i),
\\
\label{localcoeff2}
n_{2}^{i}
&=
n_{2}^{i-1}(a_i-b_i-c_i),
\\
\label{0corrcoeffiasi-1}
t_{0}^{i}
&=
t_{0}^{i-1}(a_i+b_i-c_i),
\\
\label{1corrcoeffias-1}
t_{1}^{i}
&=
t_{1}^{i-1}(a_i-b_i+c_i),
\\
\label{2corrcoeffiasi-1}
t_{2}^{i}
&=
t_{2}^{i-1}(a_i-b_i-c_i).
\end{align}
Using the above expression inductively, we get
\begin{align}
\label{localcoeff0gen}
n_{0}^{i} &=n_{0}\prod_{m=1}^{i}(a_m+b_m-c_m),\\
\label{localcoeff1gen}
n_{1}^{i} &= n_{1}\prod_{m=1}^{i}(a_m-b_m+c_m),\\
\label{localcoeff2gen}
n_{2}^{i} &= n_{2}\prod_{m=1}^{i}(a_m-b_m-c_m),\\
\label{0corrcoeffiasigen}
t_{0}^{i} &= t_{0}\prod_{m=1}^{i}(a_m+b_m-c_m),\\
\label{1corrcoeffiasgen}
t_{1}^{i}&= t_{1}\prod_{m=1}^{i}(a_m-b_m+c_m),\\
\label{2corrcoeffiasigen}
t_{2}^{i}&= t_{2}\prod_{m=1}^{i}(a_m-b_m-c_m).
\end{align}
When each Bob performs L\"uders measurement pair, we can substitute
\[
a_m=\frac{1}{4}(2+q_{0,m}+q_{1,m}),
\qquad
b_m=\frac{1}{4}(1-q_{0,m}),
\qquad
c_m=\frac{1}{4}(1-q_{1,m}),
\]
into the above equations and obtain
\begin{align}
\label{localcoeff0ss}
n_{0}^{i}
&=
\frac{n_{0}}{2^{i}}
\prod_{m=1}^{i}(1+q_{1,m}),
\\
\label{localcoeff1ss}
n_{1}^{i}
&=
\frac{n_{1}}{2^{i}}
\prod_{m=1}^{i}(1+q_{0,m}),
\\
\label{localcoeff2ss}
n_{2}^{i}
&=
\frac{n_{2}}{2^{i}}
\prod_{m=1}^{i}(q_{0,m}+q_{1,m}),
\\
\label{0corrcoeffi}
t_{0}^{i}
&=
\frac{t_{0}}{2^{i}}
\prod_{m=1}^{i}(1+q_{1,m}),
\\
\label{1corrcoeffi}
t_{1}^{i}
&=
\frac{t_{1}}{2^{i}}
\prod_{m=1}^{i}(1+q_{0,m}),
\\
\label{2corrcoeffi}
t_{2}^{i}
&=
\frac{t_{2}}{2^{i}}
\prod_{m=1}^{i}(q_{0,m}+q_{1,m}).
\end{align}

The last three expressions correspond to Eqs.~(\ref{0corrcoefffrks})--(\ref{2corrcoefffrks}) in the main text.

\section{Proof that strategies of type $(I)$ and $(II)$ are at least as effective as those of type $(III)$}
\label{suboptimalmixture}

Define
\[
f_{j,i}=\left(\frac{1+q_{j,i}}{2}\right)^2,
\]
so that the Horodecki function for the $k$th Bob is
\[
H_k
=
t_{1}^{2}\prod_{i=1}^{k}f_{0,i}
+
t_{0}^{2}\prod_{i=1}^{k}f_{1,i}.
\]
Since $|t_0|\ge |t_1|$, we have
\[
t_0^2\ge t_1^2.
\]

For each $i\leq k$, define
\begin{align}
M_i &= \max\{f_{0,i},f_{1,i}\},\\
m_i &= \min\{f_{0,i},f_{1,i}\}.
\end{align}
Consequently,
\begin{equation}
M_i\ge m_i
\end{equation}
for all $i\leq k$.

Consider an arbitrary mixed strategy of type $(III)$. Let
\begin{align}
R &= \{\,i: f_{0,i}=M_i\,\},\\
S &= \{\,i: f_{1,i}=M_i\,\},
\end{align}
so that $R\cup S=\{1,\ldots,k\}$ and $R\cap S=\varnothing$ (up to indices for which $f_{0,i}=f_{1,i}$).

Define
\begin{align}
A &= \prod_{i\in R}M_i, &
a &= \prod_{i\in R}m_i,\\
B &= \prod_{i\in S}M_i, &
b &= \prod_{i\in S}m_i.
\end{align}
Since $M_i\ge m_i$ for all $i$, it follows that
\begin{equation}
A\ge a,
\qquad
B\ge b.
\end{equation}

Using these definitions, we may write
\begin{equation}
\prod_{i=1}^{k}f_{0,i}= Ab,\quad 
\prod_{i=1}^{k}f_{1,i}=aB.
\end{equation}
Hence, the Horodecki function corresponding to the mixed strategy is
\begin{equation}
\label{mixedhorodecki}
H_k
=
t_1^2Ab+t_0^2aB.
\end{equation}

Now consider the strategy obtained by assigning the larger factor $M_i$ to each term of the product multiplying $t_{0}^{2}$ and the smaller factor $m_i$ to each term of the product multiplying $t_{1}^{2}$ for every $i\leq k$. For this strategy $M_{i}=f_{1,i}$ for all $i\leq k$. It is obtained from the mixed strategy above by applying the transformation $f_{j,i}\to f_{(j+1)\,\mathrm{mod}\,2,i}$ for $i\in R$. Therefore, we have 
\begin{equation}
\prod_{i=1}^{k}f_{0,i}= ab,\quad 
\prod_{i=1}^{k}f_{1,i}=AB.
\end{equation}
The corresponding Horodecki function is
\begin{equation}
\label{purehorodecki}
H_k'
=
t_1^2ab+t_0^2AB.
\end{equation}

Subtracting Eq.~(\ref{mixedhorodecki}) from Eq.~(\ref{purehorodecki}), we obtain
\begin{align}
H_k'-H_k
&=
t_1^2ab+t_0^2AB
-t_1^2Ab-t_0^2aB\\
&=
(A-a)(t_0^2B-t_1^2b).
\end{align}

Since $A\ge a$, we have
\[
A-a\ge0.
\]
Furthermore,
\[
t_0^2B-t_1^2b
=
(t_0^2-t_1^2)B+t_1^2(B-b)\ge0,
\]
because $t_0^2\ge t_1^2$ and $B\ge b$.

Hence,
\begin{equation}
H_k'-H_k\ge0,
\end{equation}
or equivalently,
\begin{equation}
H_k'\ge H_k.
\end{equation}

Thus, for any given mixed strategy of type $(III)$, one can construct a pure strategy of type $(I)$ or $(II)$ whose Horodecki function is at least as large. Therefore, when maximizing the Horodecki function, it is sufficient to restrict attention to pure strategies.

\section{The structure of the initial Bell diagonal quantum state suitable for unbounded unilateral nonlocality sharing}
\label{initialstate}
We begin with
\begin{equation}
\rho_{0} =\frac{1}{4}\left(\one\otimes\one +\sigma_0\otimes\sigma_0 + t_1\sigma_1\otimes\sigma_1 + t_2\sigma_2\otimes\sigma_2 \right),
\end{equation}
where
\begin{equation}
(t_0,t_1,t_2) = (1,t_1,t_2).
\end{equation}

The eigenvalues of $\rho_0$ in the Bell basis are
\begin{align}
\lambda_{\Phi^+} &= \frac{1}{4}(2-t_1+t_2), \\
\lambda_{\Phi^-} 
&= \frac{1}{4}(t_1+t_2), \\
\lambda_{\Psi^+} 
&= \frac{1}{4}(2+t_1-t_2), \\
\lambda_{\Psi^-} 
&= -\frac{1}{4}(t_1+t_2).
\end{align}

Positivity of the density operator requires
\begin{equation}
\lambda_i \geq 0.
\end{equation}

From $\lambda_{\Phi^-}\geq0$ and $\lambda_{\Psi^-}\geq0$, we obtain
\begin{equation}
\label{tconstraint}
t_1+t_2=0,
\end{equation}
so that
\begin{equation}
\label{edgeform}
(t_0,t_1,t_2)=(1,t,-t),
\qquad
-1\leq t\leq1.
\end{equation}

Thus, the allowed states form an edge of the Bell tetrahedron connecting the two maximally entangled states
\begin{equation}
(1,1,-1)
\qquad\text{and}\qquad
(1,-1,1).
\end{equation}

Equivalently,
\begin{equation}
\rho_0 = p\ket{\Psi^+}\bra{\Psi^+} +
(1-p)\ket{\Phi^+}\bra{\Phi^+},
\end{equation}
with
\begin{equation}
p \in [0,1]\ \text{and}\ t=2p-1.
\end{equation}

Hence, every Bell-diagonal state satisfying $t_0=1$ is necessarily a convex mixture of the two maximally entangled Bell states $|\Psi^+\rangle$ and $|\Phi^+\rangle$.

\end{widetext}

\end{document}